\documentclass[11pt,twoside]{article}
\usepackage[dvips]{graphics}
\usepackage{star2000}

\begin{document}

\title{The Thick Disk-Halo Interface}

\author{Gerard Gilmore}
\affil{Institute of Astronomy, Cambridge, UK}

\author{Rosemary F.G. Wyse}
\affil{Johns Hopkins University, Baltimore, MD, USA}


\begin{abstract}
The star formation history of a galaxy, explicitely here our Milky Way
Galaxy, where the most detailed information is attainable, is the
convolution of two functions. One function describes the rate of
formation of the stars which are today in the Galaxy. The second
describes the assembly of those stars into the present Galactic
potential well. There is direct evidence that this assembly continues
today, with both stars and gas being assembled into, or at least
rearranged in, the Galactic potential. But is this recent accretion
significant? Was the last significant accretion the formation
of the thick disk, some 10Gyr ago?. Recent spectroscopic studies
support this unexpected result, while dynamical studies find
increasing numbers of specific examples of smaller scale more recent
accretion. We present early results for one specific such survey, the
Anglo Australian Old Stellar populations Survey, to illustrate current
studies. 
\end{abstract}

\section{Old Stellar Populations: the Context}

While present stellar populations are a undoubtedly a manifestation of the fossil
record of Galactic evolution, quantification and interpretation of
that fossil record remains a subject of lively debate and rapid progress. 
Among the key issues are the places and times of formation of the
oldest stellar populations : the halo, thick disk and bulge - and
their overlaps and evolutionary relationships, if any. Analysis of
these stellar populations will in principle quantify the history of
merging and accretion in a typical galaxy, of great importance in
determining galactic evolution, and constraining cosmological theories
of galaxy formation.

The metallicity and kinematic distribution functions of complete
samples of long-lived stars have long been recognised as providing
unique constraints on the early stages of chemical evolution of the
Galaxy.  The main sequence lifetime of F/G dwarf stars is greater than
the age of the Galaxy and hence the chemical-abundance distribution
function of such stars provides an integrated record of the
chemical-enrichment history without the need for model-dependent
corrections for dead stars ( van den Bergh 1962; Tinsley
1980).  Pioneering studies focussed on the only
reasonably-complete sample available, which is that for stars in the
immediate solar neighborhood; in effect stars within about 30pc of the
Sun.  These samples have been sufficiently small that reliable study
of those stellar populations whose kinematics are such that member
stars spend only a small fraction of an orbit in the solar
neighborhood has necessarily been difficult. This is potentially a
serious restriction, as such stars might in principle be a major
contributor to the stellar population in a valid, representative
volume of the Galaxy.  In addition, intrinsically-rare stellar
populations are missed entirely.

Thus, it is important, in deriving a reliable determination of
Galactic structure and evolution, that one consider the joint
distributions functions over chemical abundance and kinematics.

The observational situation has been improved recently in two ways: by
collection and analysis of spectroscopic data for all-sky samples of
stars extending to somewhat greater, but still essentially local,
distances (Norris, Bessel \& Pickles 1985; Carney etal
1990; Beers etal 1999; Carney, this volume; Chiba \& Beers 2000), 
and by deeper pencil-beam surveys, to isolate {\sl in
situ} samples of old disk (Kuijken \& Gilmore 1989a), thick
disk (Gilmore, Wyse, \& Jones 1995) and halo stars. The
combination of the large but local samples with the small but distant
samples has allowed the deconvolution, to first order, of the
abundance distribution functions, and the mean velocity dispersions,
of the dominant Galactic populations. While our understanding of
Galactic structure and evolution has advanced considerably of late,
extension of these analyses has become limited by the intrinsic
breadth and overlap of the population distribution functions and by
the small size of the available {\sl in situ} samples.

The theoretical situation has also become more specific.  Though the
many dynamical, structural and chemical evolution questions one poses
concerning galactic evolution may seem well-defined and relatively
distinct, it is now clear that the answers are intimately
interrelated. For instance, galaxies probably accrete their
neighbours, so that the place of origin of a star may be far from its
present location; dynamical instabilities in disks result in the
mixing through phase space of stellar populations, further blurring
the relation between a star's present location and its birthplace. Bar
instabilities are also likely to cause significant gas transport, and
may drive star bursts and possibly nuclear non-thermal phenomena.
Major mergers may thicken disks. Bulges may be accreted, or created
during mergers.

Modern models of Galaxy formation make fairly specific predictions
concerning each of these possibilities. A detailed review is provided
by Silk \& Wyse (1993) where further
discussion may be found. For example, fashionable Cold Dark Matter
models, which contain aspects of both the monolithic (`ELS') and the
multi-fragment (`Searle-Zinn') pictures often discussed in chemical
evolution models, `predict' growth of the Galaxy about a central core,
which should contain the oldest stars. Later accretion of material
forms the outer halo and the disks, while continuing accretion will
continue to affect the kinematic structure of both the outer halo and
the thin disk.  Considerable phase-space substructure should be
detectable, when one looks sufficiently far from the Plane and the
Galactic centre that dynamical timescales are long (Ibata, Gilmore, \&
Irwin 1995; Arnold \& Gilmore 1992), and has
recently been seen with plausible significance (Helmi etal 1999).

Dissipational models for thick-disk formation predict observable
abundance gradients (cf. Burkert, Hensler and Truran 1992),
and similar scale lengths for the thick and thin disks (Ferrini
etal 1994).  Specific column-integral abundance
distributions can be calculated (numerically) for some of these models
and compared to observations.
Satellite merger models for thick disk formation require the stars
from the satellite to be detectable, as a tail in the thick disk
distribution functions below [Fe/H]=$-1$ (Silk \& Wyse, 1993)
`Continuum' models of thick disk formation from the
thin disk require that an accurately defined joint distribution
function over chemical abundance and kinematics for the oldest stars
be smooth and continuous (Norris \& Ryan, 1991).
Alternative models, such as the discrete merger model, can then be
distinguished by their prediction that the distibutions overlap in
abundance, and perhaps velocity dispersion, but not in angular
momentum (Gilmore, Wyse \& Kuijken 1989).
Most detailed models make specific predictions concerning the
abundance distribution function in a cylinder, through the Galactic
disk - the `G-dwarf problem' -- which remains widely studied, and a
valuable diagnostic of early accretion and gas flows in the disk
(Pagel \& Patchett 1975).

That is, quantitative study of the essential physics of galaxy
evolution, requires that one must study the distributions over
chemistry, kinematics and spatial structure of the oldest stars (eg
Sandage \& Fouts 1987).
Determination of the wings of the distribution functions, and their
separation or deconvolution, is however feasible, given adequate
samples.  One such project, which we introduce here, is the 
Anglo-Australian Old Stellar Populations Survey, with joint UK (the
present authors) and Australian (J Norris, K Freeman) involvement.

\section{Old Populations: what should one observe?}

 The ideal tracer of Galactic Structure is one which is selected
without any biases, does not suffer from stellar age-dependent
selection effects, is representative of the underlying populations,
and is easily observable.  Historically, the need for easy observation
restricted studies to the immediate solar neighborhood. The primary
limitation of the nearby star sample is its small size.  This
inevitably means that stars which are either intrinsically rare --
such as halo population subdwarfs -- and stars which are common but
whose spatial distribution is such that their local volume density is
small -- such as thick disk stars -- are poorly represented.  Most
recent and current efforts to extend present local volume-limited
samples to include minority populations have, for practical
observational reasons, utilized kinematically-selected samples defined
in the solar neighborhood, following the pioneering work of Eggen,
Lynden-Bell and Sandage (1962).  Subsequent correction for the
kinematic biases inherent in these samples requires careful modelling
(Norris and Ryan 1991).
An {\it in situ\/} sample, truly representative of the
dominant stellar population far from the Sun, circumvents these large,
model-dependent corrections.

Several surveys of tracer stars which can be observed {\it in situ}
are available.  Intrinsically luminous tracers are {\it a priori\/}
favored in terms of telescope time, but the likely candidates have
other characteristics that diminish their suitability: RR Lyrae stars
have intrinsic age and metallicity biases in that only stars of a
given range in metallicity and age exist in this evolutionary stage;
the accessible globular clusters are few in number; bluer horizontal
branch stars are also rare, and their color distribution depends on
chemical abundance and on the unidentified `second
parameter(s)'. K-giants are the most representative {\it evolved\/}
tracers of the spheroid, and have been used extensively.  However, one
must first identify giant stars from among the substantially larger
number of foreground K dwarfs with similar apparent magnitudes and
colors, and even with that selection, reliable determination of the
distance of a halo K giant has proven to be extremely difficult.

A desirable solution to these limitations, which has become feasible
with current multi-object spectroscopic systems and large-scale
photometric surveys, is to identify and study F/G dwarfs to
significant distances from the Sun.  This is the solution which we
have adopted.  Chemical abundances for these stars provide the
integrated record of the star formation and enrichment history during
the early stages of Galaxy formation, analogous to the local G-dwarf
distribution. Radial velocities allow discrimination between stellar
populations, when combined with abundances and spatial distributions.

Thus, the AAOSPS project is optimised to provide the next stage of
quantitative analysis of the structure, contents and evolution of the
early Milky Way, building from current observational and theoretical
expertise developed in pre-cursor phases of this study.

The scientific aim of the AAOSPS project is to determine the
distribution functions over metallicity, kinematics, and spatial
distributions of the oldest stellar populations, with particular
emphasis on the overlapping wings of each distribution function. To
achieve this, the primary technical requirement is that each of the
three variables - metallicity, radial velocity, distance - be
determined to a precision which is less than the intrinsic `cosmic'
dispersion in that parameter.

\section{The Anglo Australian Old Stellar populations Survey:AAOSPS}

Is the thick disk a merger remnant?  How does it overlap the halo?
What are the systems that merge, how frequently does this happen over
a Hubble time, and with what consequence? Could we identify stars from
the intruder, and from our own early disk?  

To address these questions, we are using the two-degree-field
multi-object spectrograph (2dF) on the Anglo-Australian Telescope,
which provides 400 spectra simultaneously, to measure the radial
velocities and chemical abundances for F/G main sequence stars at
distances from the Sun of 3--7kpc down several key lines-of-sight. 

\subsection{AAOSPS: the specific goals}

Mergers and strong interactions between galaxies happen, as evidenced
today by the Sagittarius dwarf spheroidal galaxy (Ibata, Gilmore \&
Irwin 1994, 1995; Ibata et al.  1997).  The occurrence of a `minor
merger' between the Milky Way and a small satellite galaxy provides an
attractive explanation for the thick disk (see Gilmore \& Wyse 1985;
Gilmore, Wyse \& Kuijken 1989; Freeman 1993; Majewski 1993; Walker,
Mihos \& Hernquist 1996; Huang \& Carlberg 1997; Velazquez \& White
1999).

What are the systems that merge, how frequently does this
happen over a Hubble time, and with what consequence?

Depending on the mass, density profile and orbit of the merging
satellite, `shredded-satellite' stars will leave a kinematic
signature, distinct from the canonical thick disk that will result
from the heated thin disk.  Satellites on prograde (rather than
retrograde) orbits couple better to the disk and provide more heating,
and thus are favoured to form the thick disk (e.g. Velazquez \& White
1999). Thus one might expect a signature to be visible in the mean
orbital rotational velocity of stars, and for a typical satellite
orbit, lagging the Sun by more than does the canonical thick disk.
The relative number of stars in the `shredded satellite' versus the
heated-thin disk (now the thick disk) depends on the details of the
shredding and heating processes, and is a diagnostic of them, and may
well vary strongly with location.

We are measuring radial velocities and abundances for F/G main
sequence stars at distances from the Sun of 3--7kpc.  The sample is
selected on the basis of colour and apparent magnitude (V=19-20; 0.5 <
B-V < 0.9).  The 2dF spectra have 1 Angstrom/pixel and cover
3700--4600 Angstroms. The external velocity accuracy of the data is
around 10--15 km/s, from repeat observations of program stars and from
a globular cluster standard.
We (mostly work by Norris) have extended the abundance
determination techniques developed by Beers et al. (1999).  As well as
abundances based on Ca II K, values are being determined from the
G-band of CH.  For stars with sufficient S/N we can derive
metallicities from an autocorrelation Fourier method, which uses all
the weak lines in the available spectral range.  Standard star data
show that abundances with precision better than 0.3dex are already
obtainable.  

Our primary targets are fields towards and against Galactic rotation,
to provide optimal halo/thick disk discrimination through orbital
angular momentum. We measure the time-integrated structure of the halo
and thick disk, evolved over many orbital times at these
Galactocentric distances.  This is the only statistical survey
targeting fields that probe the angular momentum, far from the thin
disk and without strong metallicity bias.

Our earlier multi-object (AUTOFIB) survey
demonstrated that there is a negligible fraction of
stars in the canonical thick disk that are younger than the globular
clusters of the same metallicity (Wyse \& Gilmore 1995) strengthening
the earlier inferences from kinematically-biased local surveys;
Gilmore \& Wyse 1985, Carney et al 1989.  This result limits the time
of the last significant merger event to be very early in the history
of the Galaxy (Gilmore, Wyse \& Jones 1995), challenging standard CDM
cosmologies (cf Wyse 2001).

\begin{figure}[!ht]
\plotfiddle{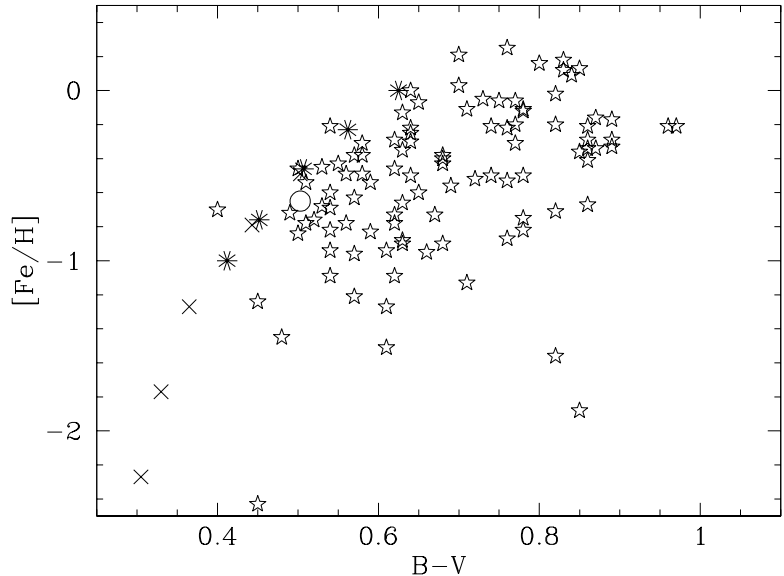}{3.25in}{270}{40}{40}{-155}{240}
\caption{Scatter plot of iron abundance {\it vs\/} B-V colour for
thick disk F/G stars, selected {\it in situ\/} in the South Galactic Pole 
at 1-2kpc above the
Galactic Plane (stars), together with the 14~Gyr turnoff colours
(crosses) from VandenBerg \& Bell (1985; Y=0.2) and 15~Gyr turnoff colours
(asterisks) from VandenBerg (1985; Y=0.25).  The open circle represents the
turnoff colour (de-reddened) and metallicity of 47~Tuc (Hesser et
al.~1987).  The vast majority of thick disk stars lie to the red of
these turnoff points, indicating that few, if any, stars in this
population are younger than this globular cluster.  This figure is
from Wyse (2001).}
\end{figure}

\subsubsection{First Results:}

We have detected a substantial population of low metallicity stars
with disk-like kinematics, intermediate between those of the canonical
thick disk and the canonical (non-rotating) stellar halo.  Figure 1
shows the radial velocity histogram for around 900 stars with high
signal-to-noise spectra, in a line-of-sight for which, at these
distances, radial V-velocity is approximately 0.8 times the rotational
lag behind the Sun's orbit. Our efficient selection against thin disk
contamination -- radial velocity near zero -- is apparent.
The smooth Gaussian represents a smooth halo in this
line-of-sight, appropriately normalised to fit the high-velocity data.
The canonical thick disk has a rotational lag of some 40km/s.  Thus
the large number of stars with radial velocity around 100km/s is not
expected, and probably traces a new kinematic component of the Milky
Way Galaxy.

\begin{figure}[htb!]
\plotfiddle{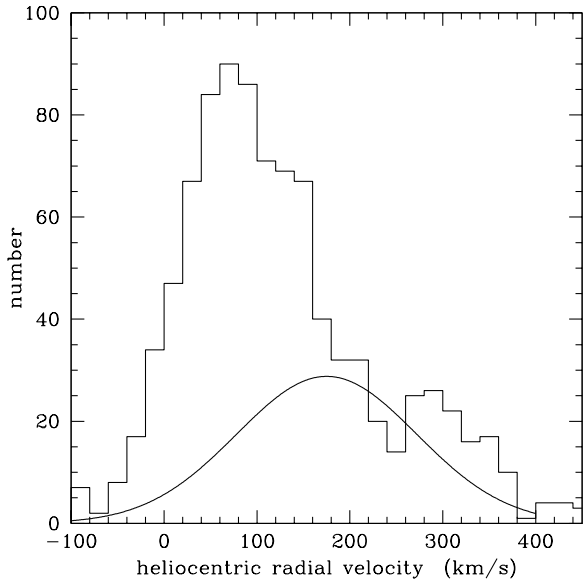}{7.5cm}{0}{40}{40}{-120}{-60}
\caption{Radial velocity histogram for around 900 stars in a
line-of-sight chosen to probe orbital rotation.  The smooth curve
represents the stellar halo, and while it clearly is a reasonable
description of the shape of the distribution of the 
highest velocity stars, it
fails to describe the majority of the stars.  The canonical thick disk
provides the stars with radial velocity of less than 100km/s; the
broad shoulder between 100km/s and 200km/s is not expected. }
\end{figure}

The 100km/s stars are best interpreted as being the actual debris of the
satellite. The debris `stream' is  detected in
widely-separated lines-of-sight, but requires a larger
statistically-significant sample for confirmation and to allow
quantification of the properties of the former satellite galaxy.  This
quantification would be a strong constraint on hierarchical models of
galaxy formation.

\begin{figure}[htb!]
\begin{center}
\plotfiddle{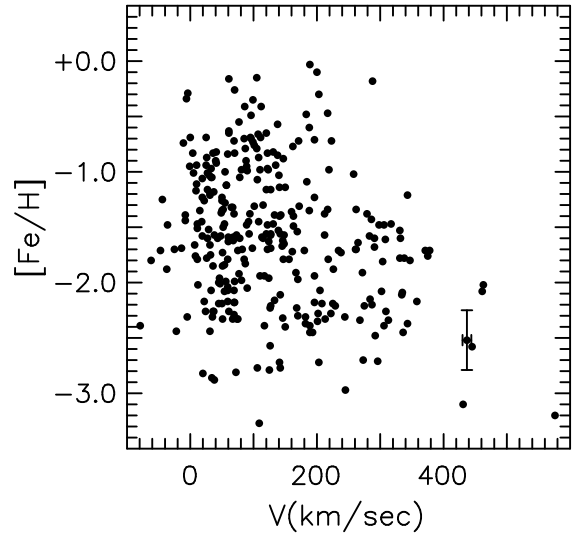}{7.5cm}{0}{100}{100}{-270}{-180}
\caption{2dF metallicities vs heliocentric radial velocity, for our
stars with sufficient S/N, in one line of sight.  Radial V velocity at
these distances in this line-of-sight is approximately Galactic
rotational velocity, with 180km/s approximately zero net orbital
rotation. The many stars with low metallicity and disk-like (small-V)
velocities are the identification of a low metallicity tail of the
thick disk. There is no metal-rich halo.  Lumpiness in the figure is
apparent: if real it will indicate phase-space structure, allowing
quantification of the past merger history.}
\end{center}
\end{figure}

\subsubsection{Metallicity and Phase Space Structure?}

The combination of kinematics and metallicity provides the best
constraints on stellar populations.  We are quantifying the
distributions of the Galaxy's populations in metallicity-velocity
space, to higher precision than simple Gaussian fits: it is these
distributions which encode galaxy formation.

Our first results (figure 2) show substantial numbers of stars with
very low metallicities, and very high angular momentum: that is, we
have discovered the much sought metal-weak thick disk, perhaps the
remnants of the Milky Way's last big merger.

Our velocity accuracy is adequate to identify any high-frequency
phase-space structure which may exist.  The clumpiness in the
metallicity vs velocity diagram shown, and the spikiness in the number
vs Galactic rotation velocity data, are kinematically resolved: our
goal now is to obtain sufficient numbers of adequate quality spectra to
quantify the statistical significance of these features, and the scale
length on the sky with which they are associated.

The key result is apparent from figure 2.  At every scale
we see mildly significant structure, and deviations from Gaussians:
are the groupings of stars in phase space, and the deviations from
kinematic smoothness, physical?  Statistical tests show that
substructure is significant, but only marginally, and only when the
data are restricted by angular scale length on the sky. This is
just what some spaghetti models predict (eg. Helmi \& White 1999; Helmi
et al 1999; Harding et al 2001). We are continuing to investigate its
reality.

\section{Conclusions:}

Modern large area surveys, complemented by deeper multi-object
facilities, are obtaining and analysing the combination of metallicity
and kinematic data for large samples of Galactic stars in the thick
disk -- halo interface.  Our preliminary results for one such study,
AAOSPS, show intriguing, but low statistical significance, deviations
from canonical distributions.  These may signal the remnant of the
satellite whose merger with the young Milky Way formed the thick disk,
the last high-impact merger that our Galaxy experienced.  Further, the
amplitude of small-scale lumpiness constrains the more recent merger
history, with recent detections from HIPPARCOS local data again being
complemented by more distant studies.  We are on the way to
deciphering the fossil record of the physical processes in the
formation and evolution of a typical large disk galaxy, the Milky Way.

\end{document}